\documentclass{sig-alternate-10pt}

\usepackage{array}
\usepackage{graphicx}
\usepackage{float}
\usepackage{stfloats}
\usepackage{lipsum}
\usepackage{tabularx}
\usepackage[tight,footnotesize]{subfigure}
\usepackage{caption}
\usepackage{epstopdf}
\usepackage{amsmath}
\usepackage{amssymb}
\usepackage{paralist}
\usepackage{mathrsfs}
\usepackage{url}
\usepackage{mathbbol}
\usepackage{times}

\usepackage{booktabs}
\usepackage{verbatim}

\usepackage{bm}			
\usepackage{cite}

\usepackage{amsthm}

\usepackage[table]{xcolor}
\usepackage{color,colortbl}

\definecolor{Gray}{gray}{0.9}

 \usepackage[normalem]{ulem}

\begin{document}


%
%
%

%


\title{Monitor, Detect, Mitigate: Combating BGP Prefix Hijacking in Real-Time with ARTEMIS}

\author{
Pavlos Sermpezis, Gavriil Chaviaras, Petros Gigis, and Xenofontas Dimitropoulos\\
\affaddr{FORTH, Greece}\\
\email{\{sermpezis, gchaviaras, gkigkis, fontas\}@ics.forth.gr}
}

\maketitle

\begin{abstract}

The Border Gateway Protocol (BGP) is globally used by Autonomous Systems (ASes) to establish route paths for IP prefixes in the Internet. Due to the lack of authentication in BGP, an AS can \textit{hijack} IP prefixes owned by other ASes (i.e., announce illegitimate route paths), impacting thus the Internet routing system and economy. To this end, a number of hijacking detection systems have been proposed. However, existing systems are usually third party services that -inherently- introduce a significant delay between the hijacking detection (by the service) and its mitigation (by the network administrators). To overcome this shortcoming, in this paper, we propose ARTEMIS, a tool that enables an AS to \textit{timely} detect hijacks on its own prefixes, and \textit{automatically} proceed to mitigation actions. To evaluate the performance of ARTEMIS, we conduct real hijacking experiments. To our best knowledge, it is the first time that a hijacking detection/mitigation system is evaluated through extensive experiments in the real Internet. Our results (a) show that ARTEMIS can detect (mitigate) a hijack within a few seconds (minutes) after it has been launched, and (b) demonstrate the efficiency of the different control-plane sources used by ARTEMIS, towards monitoring routing changes.

\end{abstract}

\section{Introduction}\label{sec:intro}
The inter-domain routing in the Internet takes place over the -globally adopted- Border Gateway Protocol (BGP)~\cite{BGPv4}. Autonomous Systems (ASes) use BGP to advertise routing paths for IP prefixes to their neighboring ASes. Since BGP is a distributed protocol and authentication of advertised routes is not always feasible, it is possible for an AS to advertise illegitimate route paths for IP prefixes. These paths can propagate and ``infect" many ASes, or even the entire Internet, impacting thus severely the Internet routing system and/or economy~\cite{hijack-YouTube,hijack-ChinaTelecom,hijack-BitCoins,Ramachandran-BGP-spammers-CCR-2006, Vervier-mind-blocks-NDSS-2015}.

This phenomenon, called \textit{BGP prefix hijacking}, is frequently observed~\cite{Vervier-mind-blocks-NDSS-2015}, and usually caused by router misconfigurations~\cite{hijack-YouTube,hijack-ChinaTelecom} or malicious attacks~\cite{hijack-BitCoins,Ramachandran-BGP-spammers-CCR-2006, Vervier-mind-blocks-NDSS-2015}. Some examples of real BGP hijacking cases include: (a) a Pakistan's ISP in 2008, due to a misconfiguration, hijacked the YouTube's prefixes and disrupted its services for more than $2$ hours~\cite{hijack-YouTube}; (b) China Telecom mistakenly announced $\sim37000$ IP prefixes (corresponding to $15\%$ of the whole BGP table) in 2010, causing routing problems in the Internet~\cite{hijack-ChinaTelecom}; and (c) hackers performed several hijacking attacks, through a Canadian ISP, to redirect traffic and steal thousands dollars worth of bitcoins in 2014~\cite{hijack-BitCoins}.

To prevent prefix hijackings, several proactive mechanisms for enhancing the BGP security have been proposed~\cite{Kent-secure-BGP-JSAC-2000,Subramanian-listen-whisper-NSDI-2004,bgpsec-specification-2015,rpki-rfc,Karlin-PGBGP-ICNP-2006}. These mechanisms need to be globally deployed to be effective. However, despite the standardization efforts~\cite{bgpsec-specification-2015,rpki-rfc}, their deployment is slow due to political, technical, and economic challenges, leaving thus the Internet vulnerable to BGP hijacks. 

Therefore, currently, reactive mechanisms are used for defending against prefix hijackings: after a hijacking is detected, network administrators are notified (e.g, through mailing lists~\cite{Shi-Argus-IMC-2012}, or dedicated services~\cite{Lad-Phas-Usenix-2006}), in order to proceed to manual actions towards its mitigation (e.g., reconfigure routers, or contact other ASes to filter announcements). A number of systems have been proposed for detecting prefix hijacking, based on control plane (i.e., BGP data) and/or data plane (i.e,. pings/traceroutes) information~\cite{Chi-cyclops-CCR-2008,Lad-Phas-Usenix-2006,bgpmon,Zhang-Ispy-CCR-2008,Zheng-lightweight-hijacks-2007,Shi-Argus-IMC-2012,Hu-accurate-hijacks-SP-2007}. Most of them, are designed to operate as third-party services (external to an AS) that monitor the Internet, and upon the detection of a suspicious incident, notify the involved ASes. Although this approach has been shown to be able to detect suspicious routing events in many cases, two main issues still remain unsolved: (i) the detection might not be accurate, since the suspicious routing events might not correspond to hijacks, but be caused by, e.g., traffic engineering; and (ii) the mitigation is not automated, increasing thus significantly the time needed to resolve a hijack.

In this paper, we propose a reactive mechanism/system, which we call ARTEMIS (\textit{Automatic and Real-Time dEtection and MItigation System})\footnote{A demo of ARTEMIS is to appear in ACM Sigcomm 2016~\cite{ARTEMIS-Demo-Sigcomm-2016}.}, that aims to be operated by an AS itself, rather than a third-party, to timely detect and mitigate hijackings against its \textit{own} prefixes in an automatic way. ARTEMIS (i) exploits the most recent advances in control-plane monitoring to detect in near real-time prefix hijackings, and (ii) immediately proceeds to their automatic mitigation (Section~\ref{sec:artemis}).

We then conduct several real hijacking experiments in the Internet using the PEERING testbed and analyze the effect of various network parameters (like, type of hijacking, hijacker / defender-AS location and connectivity) on the performance of ARTEMIS. We show that it is possible to detect and mitigate prefix hijacking within {\it few seconds} from the moment the offending announcement is first made. This is a major improvement compared to present approaches, which require slow procedures, like manual verification and coordination. The timely mitigation of ARTEMIS, prevents a hijacking from spreading to just, e.g., 20\%-50\% of the ASes that would be affected otherwise (Section~\ref{sec:evaluation}).

Finally, we discuss related work in hijacking detection systems and measurement studies, and compare it to our study, in order to highlight the new capabilities that are offered with ARTEMIS (Section~\ref{sec:related}). We conclude our paper by discussing the potential for future applications and extensions of ARTEMIS (Section~\ref{sec:conclusion}).

\section{ARTEMIS}\label{sec:artemis}

In this section, we first present the different sources that are used by ARTEMIS for control-plane \textit{monitoring} (Section~\ref{sec:sources}), and then describe the \textit{detection} (Section~\ref{sec:detection}) and \textit{mitigation} (Section~\ref{sec:mitigation}) services.

\subsection{Control-Plane Data Sources}\label{sec:sources}

For the monitoring service, ARTEMIS combines multiple control-plane sources to (a) accelerate the detection of a hijacking (i.e., minimum time of all sources), and (b) have a more complete view of the Internet (i.e., from the vantage points of all the sources). ARTEMIS receives control-plane information from publicly available sources, namely, the BGPmon tool~\cite{bgpmon}, the live-streaming service of RIPE-RIS~\cite{ripe-ris-real-time}, and the Periscope platform~\cite{periscope}.

\underline{Remark:} ARTEMIS supports the BGPstream tool~\cite{bgpstream} as well. However, during our experiments, the {BGPstream} service was unavailable, and, thus, we do not use it in this~paper.

In the following, we present the main features of these control-plane sources.

\textbf{BGPmon~\cite{bgpmon}} is a tool that monitors BGP routing information in real-time. It is connected to, and collects BGP updates and routing tables (RIBs) from BGP routers of: (a) the RouteViews sites and (b) a few dozen of peers around the world; at the time we conducted our study, BGPmon had $43$ vantage points, in total. BGPmon provides the live BGP data, as an XML stream.

\textbf{RIPE RIS streaming service~\cite{ripe-ris-real-time}.} The RIPE's Routing Information System (RIS) is connected to route collectors (RCs) in several locations around the world, and collects BGP updates and RIBs. In the standard RIPE RIS~\cite{ripe-ris}, the data can be accessed via the raw files (in MRT format) or RIPEstat. The delay for BGP updates is $\sim 5min$ and $\sim 8h$ for RIBs. Recently, RIPE RIS offers a streaming service~\cite{ripe-ris-real-time} that provides live information from $4$ RCs via websockets. The live streaming service of RIPE RIS, which we use in ARTEMIS, has currently $3$ RCs in Europe and $1$ RC in Africa; all of them are located in large IXPs.

\textbf{Periscope~\cite{periscope}} is a platform that provides a common interface for issuing measurements from Looking Glass (LG) servers. Through Periscope, a user can send a command to a number of chosen LGs to ask for control-plane (\textit{show ip bgp}) or data-plane (\textit{traceroute/ping}) information. The status and the output of the measurements can be retrieved at any time (even before its completion). Periscope currently provides access to $1691$ LG servers.

\textbf{BGPstream~\cite{bgpstream}} is an open-source framework for live (and historical) BGP data analysis. It enables users to quickly inspect raw BGP data from the command-line, or through a Python and C/C++ API. BGPstream provides live access to RouteViews and RIPE RIS data archives. While the delay of acquiring the data from these two services is considerable ($5$min and $15$min, respectively, for BGP updates), BGPstream recently introduced a service for live access to a stream of BGP data from BMP-enabled RouteViews collectors (with only $\sim 1$min delay). In total, BGPstream receives data from $76$ route collectors, from all its providers.

\subsection{Prefix Hijacking Detection with ARTEMIS}\label{sec:detection}
The detection service of ARTEMIS aims to detect hijacks in (i) \textit{real-time} and (ii) \textit{without false positives}, while monitoring the (iii) \textit{entire Internet} in a (iv) \textit{light-weight} fashion.

The detection service continuously receives from the $3$ control-plane sources (see Section~\ref{sec:sources}) information about the BGP route paths for the monitored prefixes, as they are seen at the different vantage points (e.g., route collectors, LG servers). This routing information is compared with a local file that defines the legitimate origin-ASNs for each IP prefix that is owned by the operator of ARTEMIS; any violation denotes a hijacking. Since operator has full knowledge on the legitimate origin-ASNs for its  prefixes, the detection service returns \textit{no false positives}.

With the combination of $3$ sources, the detection can take place when an illegitimate route path is received by any of the sources. This is always faster than using only one source, and can decrease the time needed for detection. Using multiple sources gives also the possibility to benefit from the \textit{large number of vantage points} they have around the globe. This is important, because a hijacking might affect only a part of the Internet, due to BGP policies and shortest-path routing~\cite{Shi-Argus-IMC-2012,Lad2-Understanding-resiliency-hijacks-DSN-2007}.

Finally, ARTEMIS aims to impose limited load on the used third-party services, so that potentially 100s ASes (that run ARTEMIS) could use them in parallel. ARTEMIS needs to receive only the data (i.e., the part of the BGP tables, or specific BGP updates) that correspond to the local prefixes. As a result, \textit{the imposed load is low}, since (a) BGPmon and RIPE RIS (as well as, BGPstream) are services/tools designed and optimized to provide streams of live data to many users simultaneously, (b) and Periscope has already a limit in the rate of requests to avoid overloading of LG servers. Similarly, \textit{the consumption of network resources is very low}, allowing thus a single AS to monitor many prefixes.

\subsection{Automatic Prefix Hijacking Mitigation}\label{sec:mitigation}
The goals of a mitigation mechanism are to be (i) \textit{fast} and (ii) \textit{efficient}, and (iii) have \textit{low impact} on the Internet routing system. 

Currently mitigation relies on manual actions, e.g., after a network administrator is notified for a prefix hijacking, she proceeds to reconfiguration of the BGP routers, or contacts other administrators to filter the hijacker's announcements. As it becomes evident, this manual intervention introduces a significant delay (e.g., in the YouTube hijacking incident in 2008~\cite{hijack-YouTube}, a couple of hours were needed for the mitigation of the problem). Hence, our primary focus  is to accelerate the mitigation. To this end, we implement an automatic mitigation mechanism, which starts the mitigation \textit{immediately} after the detection, i.e., without manual intervention.

Specifically, when ARTEMIS detects a hijacking in a prefix, let \textit{10.0.0.0/23}, it proceeds to its de-aggregation: it sends a command to the BGP routers of the AS to announce the two more-specific prefixes, i.e., \textit{10.0.0.0/24} and \textit{10.0.1.0/24}. The sub-prefixes will disseminate in the Internet and re-establish legitimate route paths, since more specific prefixes are preferred by BGP. Prefix de-aggregation, as described above, is efficient for \textit{/23} or less specific (i.e., \textit{/22, /21, ...}) prefixes. However, when it comes to hijacking of \textit{/24} prefixes, the de-aggregation might not be always efficient, since prefixes more specific than \textit{/24} are filtered by most routers~\cite{Bush-Internet-optometry-IMC-2009}. Although this is a shortcoming of the de-aggregation mechanism, it is not possible to overcome it in an automatic way (manual actions are needed); to our best knowledge, only solutions that require the cooperation of more than one ASes could be applied~\cite{Zhang-practical-defenses-CoNext-2007,Kotronis-Routing-Centralization-ComNets-2015}.

The de-aggregation mechanism of ARTEMIS, increments the number of entries in the BGP routing table by $1$ per hijacked prefix. However, since the number of concurrent hijackings is not expected to be large, and the duration of a hijacking is limited, the imposed overhead is low. 

Finally, since ARTEMIS monitors continuously the control-plane of the Internet, from many vantage points, it becomes possible to monitor in real-time the process of the mitigation. This enables a network administrator to see how efficient the mitigation is, and if needed to proceed to further (e.g., manual) actions or to rely exclusively on the de-aggregation mechanism.

\section{Evaluation with a real AS}\label{sec:evaluation}
In this section, we conduct experiments in the  Internet, to investigate (a) the overall performance of ARTEMIS, and  (b) the efficiency of the different sources presented in Section~\ref{sec:sources} for monitoring the control-plane of the Internet. In Section~\ref{sec:experimental-setup} we provide the details for the setup of our experiments, and present the results in Section~\ref{sec:experiments-results}.

\subsection{Experimental Setup}\label{sec:experimental-setup}
In our experiments, we conduct \textit{real} hijackings  in the Internet. We use the PEERING testbed~\cite{Schlinker-PEERING-HotNets-2014,peering-website}, which provides the possibility to announce IP prefixes from real ASNs to the Internet; both the IP prefixes and the ASNs are owned by PEERING, hence, our experiments have no impact on the connectivity of other ASes.

Specifically, we create a virtual AS in PEERING, and connect it to one or more real networks. This AS (which we call ``legitimate'' AS) announces an IP prefix and uses ARTEMIS to continuously monitor this prefix. We also create another virtual AS (the ``hijacker'' AS) in PEERING, connect it to a real network in a different location, and hijack the prefix of the legitimate AS.

\subsubsection{The PEERING testbed}
PEERING is a testbed that enables researchers to interact with the Internet's routing system. It connects with several real networks at universities and Internet exchange points around the world. The users of PEERING can announce IP prefixes using multiple ASNs owned by PEERING as the origin-AS.

In our experiments, we use the connections of PEERING to three real networks/sites (Table~\ref{table:peering-sites})\footnote{PEERING peers with $88$ organizations in AMS-IX~\cite{peering-website}. Statistics for the number providers, customers, and peers for each AS are from~\cite{as-rank-website}.}. We are given authorization to announce the prefix \textit{184.164.228.0/23} (as well as, its sub-prefixes), and use the AS numbers \textit{61574} (for the legitimate AS) and \textit{61575} (for the hijacker AS).

\begin{table*}
\centering
\caption{PEERING sites}\label{table:peering-sites}
\begin{tabular}{c|cccccc}
{}				&{Organization}							&{Location}				&{ASN}	&{\#providers}	&{\#customers}	&{\#peers}\\
\hline
{\textbf{AMS}}	&{Amsterdam Internet Exchange (AMS-IX)}	&{Amsterdam, NL}			&{1200}	&{-}				&{-}				&{509}\\
{\textbf{ISI}}	&{Los Nettos Regional Network}			&{Los Angeles (CA), US}	&{226}	&{4}				&{19}			&{19}\\
{\textbf{GAT}}	&{Georgia Institute of Technology}		&{Atlanta (GA), US}		&{2637}	&{4}				&{1}				&{6}\\
\end{tabular}
\end{table*}

\subsubsection{Types of prefix hijacking attack}
We test ARTEMIS in two different types of hijacking attacks: (a) exact prefix hijacks, and (b) sub-prefix hijacks.

\textbf{Exact prefix hijacking} is a common attack type where the hijacker announces the same prefix that is announced by the legitimate AS. Since shortest route paths are typically preferred, only \textit{a part of the Internet} that is closer to the hijacker (in number of AS-hops) switches to route paths towards the hijacker. Exact prefix hijacks typically infect a few tens or hundreds of ASes~\cite{Shi-Argus-IMC-2012}, from  small stub networks to large tier-1 ISPs~\cite{Lad2-Understanding-resiliency-hijacks-DSN-2007}. In our experiments, the legitimate AS announces the prefix \textit{184.164.228.0/23}; then the hijacker announces the same prefix. To mitigate the attack, the legitimate AS, announces the sub-prefixes \textit{184.164.228.0/24} and \textit{184.164.229.0/24}.

\textbf{Sub-prefix hijacking} contributes around 10\% of all stable hijackings in the Internet~\cite{Shi-Argus-IMC-2012}. The hijacker announces a more specific prefix, which is covered by the prefix of the legitimate AS. Since in BGP more specific prefixes are preferred, \textit{the entire Internet} switches to routing towards the hijacker for the announced sub-prefix. We configure ARTEMIS to monitor the \textit{184.164.228.0/22} prefix\footnote{Since we have access only to the \textit{/23} prefix, we do not announce the \textit{/22} prefix; we only assume it is announced. However, this does not affect the outcome of the experiments.}. The hijacker announces the prefix \textit{184.164.228.0/23}. The attack is mitigated by de-aggregating the hijacked prefix, i.e., the legitimate AS announces the two \textit{/24} prefixes.

\subsubsection{Experiments}
The experiment process comprises the following steps:

\noindent\textbf{(1)} The legitimate AS (\textit{AS61574}) announces the IP prefix, and we wait $20$min for BGP convergence.

\noindent\textbf{(2)} The hijacker AS (\textit{AS61575}) announces the IP prefix (or, sub-prefix).

\noindent\textbf{(3)} ARTEMIS detects the hijacking.

\noindent\textbf{(4)} ARTEMIS starts the mitigation, i.e., the legitimate AS announces the de-aggregated sub-prefixes. 

\noindent\textbf{(5)} We monitor the mitigation process for $30$min, and end the experiment by withdrawing all announcements.

We conduct experiments for a number of different scenarios, varying the (a) \textit{location/site} of the legitimate and hijacker ASes, and (b) \textit{number of upstream providers} of the legitimate AS. We repeat each scenario/experiment $10$ times. The experiments took place in May-June 2016.

\underline{Remark:} While normally ARTEMIS proceeds immediately after a hijacking detection to its mitigation, in some experiments we add a $30$min delay between steps 3 and 4, i.e., we defer the mitigation. This allows us to investigate the efficiency of the different control-plane sources, i.e., how much time each of them needs to detect the hijacking.

\subsubsection{Configuration of the control-plane sources}
\textit{BGPmon} provides to ARTEMIS a stream of all the updates it receives from its peers. Hence, configuration is not needed; filtering and detection are internal services of ARTEMIS. 

\textit{RIPE RIS} needs only the information about the monitored prefix, and returns to ARTEMIS only the BGP messages that correspond to announcements for this prefix.

In \textit{Periscope}, due to the limit on the rate of measurements per user, only a subset of the total 1691 LG servers can be used. To conform to the rate-limit, we use $18$ LG servers, which we select based on their performance (response time, availability) and location. The selected set consists of $11$~LGs in Europe, $2$ in Asia, $4$ in North America, and $1$ in Australia.

\subsection{Results}\label{sec:experiments-results}

\subsubsection{Performance of control-plane sources}
\begin{figure}
\subfigure[detection delay]{\includegraphics[width=0.49\linewidth]{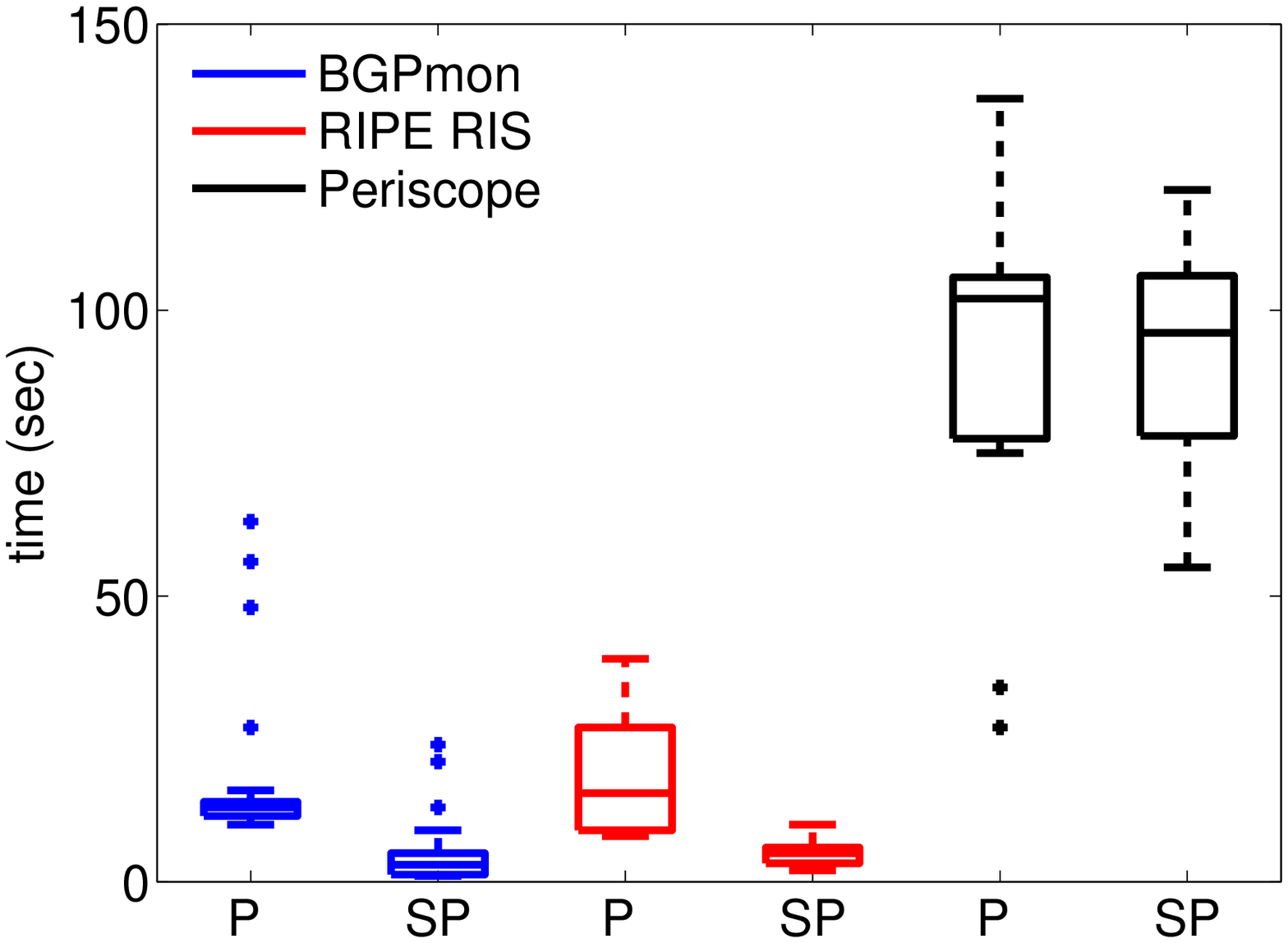}\label{fig:boxplots-detection-delay-per-tool}}
\subfigure[mitigation delay]{\includegraphics[width=0.49\linewidth]{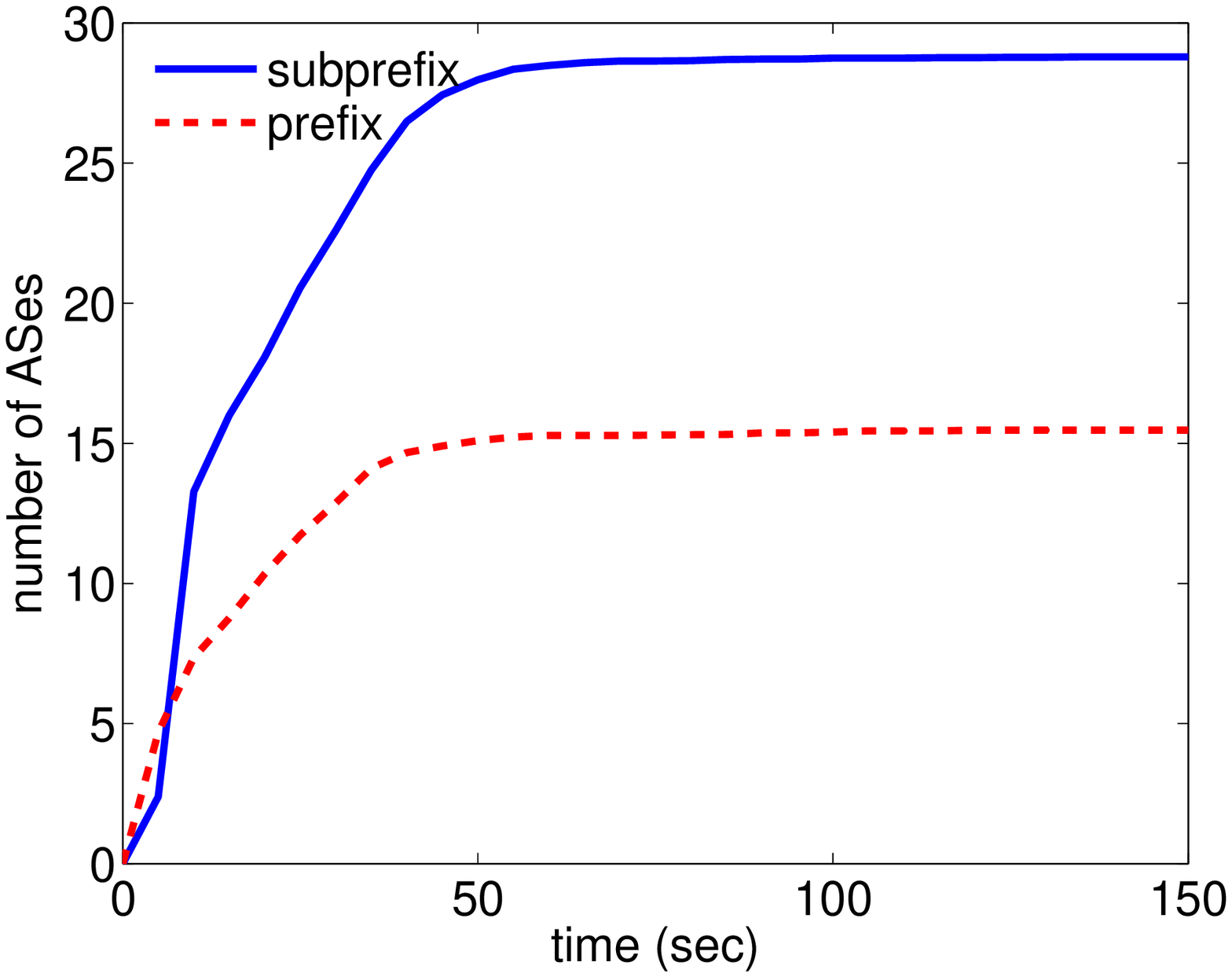}\label{fig:mitigation-AS-vs-time}}
\caption{(a) Boxplots of the detection delay of the different control-plane sources among all the prefix (P) and sub-prefix (SP) hijacking experiments. (b) Number of ASes (with a vantage point) that switched back to the legitimate AS (y-axis) vs. the time the mitigation has been launched (x-axis).}
\label{fig:detection-mititgation}
\end{figure}

The performance of ARTEMIS depends on the control-plane sources it uses. Therefore, to obtain an initial understanding about the capabilities and limitations of ARTEMIS, we present in Fig.~\ref{fig:detection-mititgation} experimental results that demonstrate the efficiency and characteristics of the different control-plane sources in hijacking detection (Fig.~\ref{fig:boxplots-detection-delay-per-tool}) and mitigation monitoring (Fig.~\ref{fig:mitigation-AS-vs-time}). 

Fig.~\ref{fig:boxplots-detection-delay-per-tool} shows how much time is needed by BGPmon, RIPE RIS, and Periscope to observe an illegitimate route, after it has been announced from the hijacker AS. We present the distribution of the times (among different experiments) for both attack types: prefix and sub-prefix hijacking. 

A first observation is that the streaming services (BGPmon and RIPE RIS) observe the hijack in $\leq1$min in most cases, and are significantly faster than Periscope ($1$-$2$min), which monitors the control-plane by periodically issuing measurements from LG servers. This is due to the response delay of the LGs, as well as, a limit in the minimum time interval between consequent measurements imposed by Periscope. 

The detection delay in the sub-prefix attack case (SP) is -on average- lower than in prefix hijacking (P). This is because a sub-prefix hijacking appears in the whole Internet, whereas prefix hijacking affects only a fraction of it. This partial infection of the Internet can be faster observed by BGPmon that has more vantage points than RIPE RIS, as is indicated by the lower mean value and variance of BGPmon in the prefix hijacking case.

In Fig.~\ref{fig:mitigation-AS-vs-time}, we show the mitigation progress as it has been observed by the ASes with a vantage point, i.e., an RC feed or an LG server, in all sources. The average number of ASes that have been infected by the hijacker and switched back to the legitimate routes, are $29$ and $15$ in the SP and P case, respectively. Despite the differences, both attacks can be quickly mitigated; $45\%$ (SP) and $50\%$ (P) of the ASes re-establish legitimate routes in $10$sec after the mitigation was launched, while almost complete mitigation is achieved in less than $1$min. Furthermore, Figs.~\ref{fig:boxplots-detection-delay-per-tool} and~\ref{fig:mitigation-AS-vs-time} hint to an interesting trade-off: more vantage points (and, thus, ASes) can be monitored by Periscope, however, this comes with an increase in the detection delay compared to BGPmon and RIPE RIS.

\subsubsection{Effect of network connectivity}
We now proceed to test the efficiency of ARTEMIS under various scenarios of network connectivity. Fig.~\ref{fig:detection-location} illustrates the effect of the (i) hijacker site and (ii) number of upstream providers of the legitimate AS\footnote{Our results do not significantly variate with the location of the legitimate AS or the number of upstream providers of the hijacker.}. In the prefix hijacking case (Fig.~\ref{fig:boxplots-delay-vs-location-prefix}), when the hijacking is triggered by a well connected site, as in the case of AMS that it peers with $88$ real networks, the detection of the hijacking can be done in around $10$sec. When the connectivity of the hijacker AS is low, as in the GAT case that there are less than a dozen of directly connected networks, the detection delay is always higher than $15$sec and can need up to $1$min (the average detection delay is around $30$sec). These findings are intuitive and consistent with the conclusions of the simulation study in~\cite{Lad2-Understanding-resiliency-hijacks-DSN-2007}; adding to this, they quantify for the first time the effects of the hijacker's connectivity with real experiments.

In Fig.~\ref{fig:boxplots-delay-vs-location-prefix}, we can also observe that when the connectivity of the legitimate AS increases, i.e., $2$ upstream providers, the detection delay (slightly) increases as well. This is due to the fact that with $2$ upstream providers, more ASes are closer to the legitimate AS (in terms of AS-hops) than the hijacker, and thus the effect of prefix hijacking is lower (and, consequently, its detection becomes more difficult).

In contrast to the prefix hijacking case, when the hijacker announces a sub-prefix (Fig.~\ref{fig:mboxplots-delay-vs-location-subprefix}), the connectivity of the involved networks does not play a crucial role. The effect of the hijacking is large and the detection is always completed within $10$sec, and on average it needs only $3$sec!

\underline{Remark:} In~\cite{Shi-Argus-IMC-2012} it is shown that the ``detection delay'' of Argus (a state-of-the-art hijacking detection system) is less than $10$sec for $>60\%$ hijacks. However, this delay, let $T_{dd}$, refers to the time needed to infer that an observation of an illegitimate route corresponds to a hijacking attack; i.e., if Argus uses the same control-plane sources as ARTEMIS, the total detection delay of Argus is $T_{Argus} = T_{ARTEMIS}+T_{dd}\geq T_{ARTEMIS}$.

\subsubsection{Gains of automatic mitigation}
\begin{figure}
\subfigure[prefix hijacking]{\includegraphics[width=0.49\linewidth]{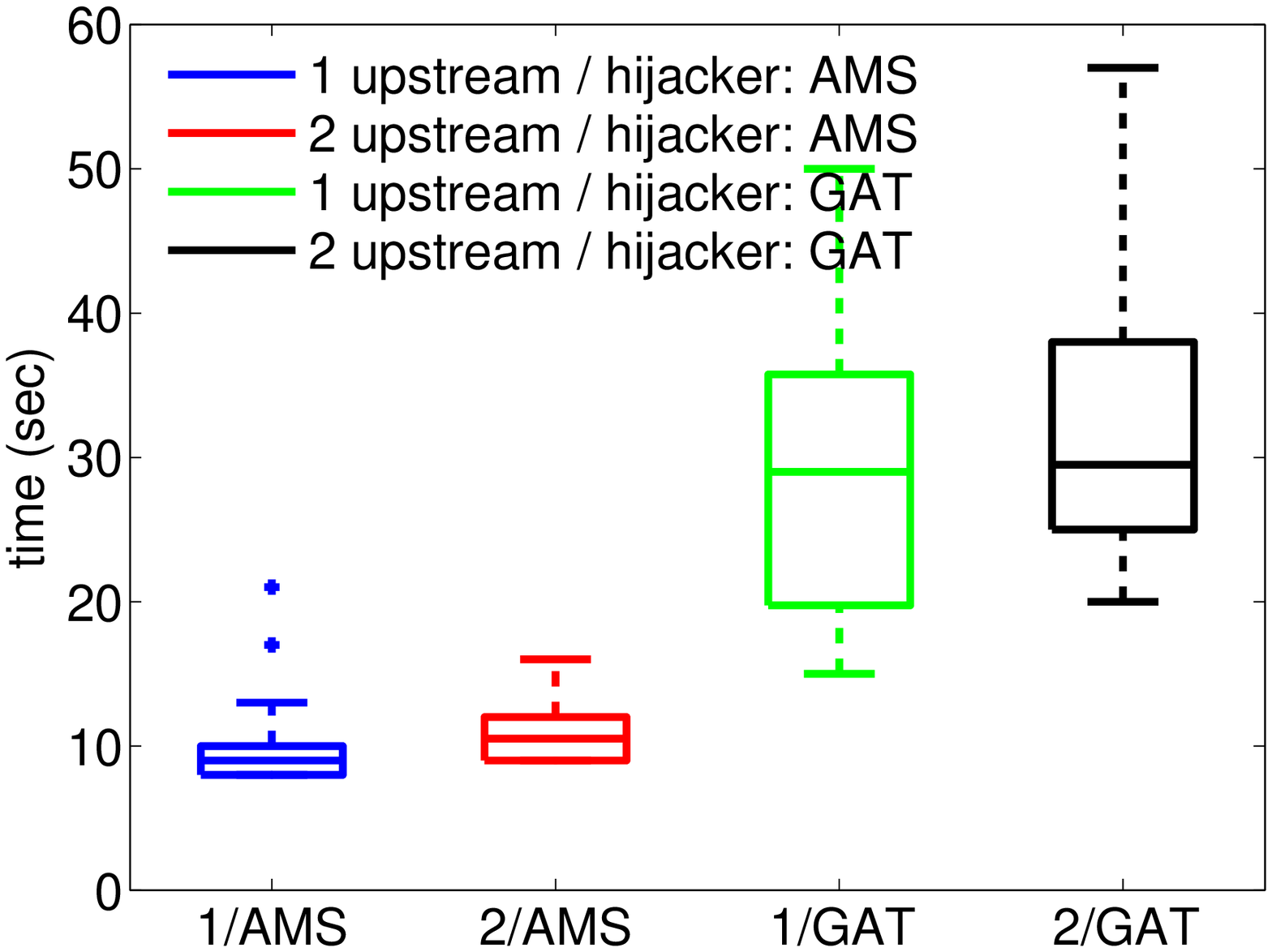}\label{fig:boxplots-delay-vs-location-prefix}}
\subfigure[sub-prefix hijacking]{\includegraphics[width=0.49\linewidth]{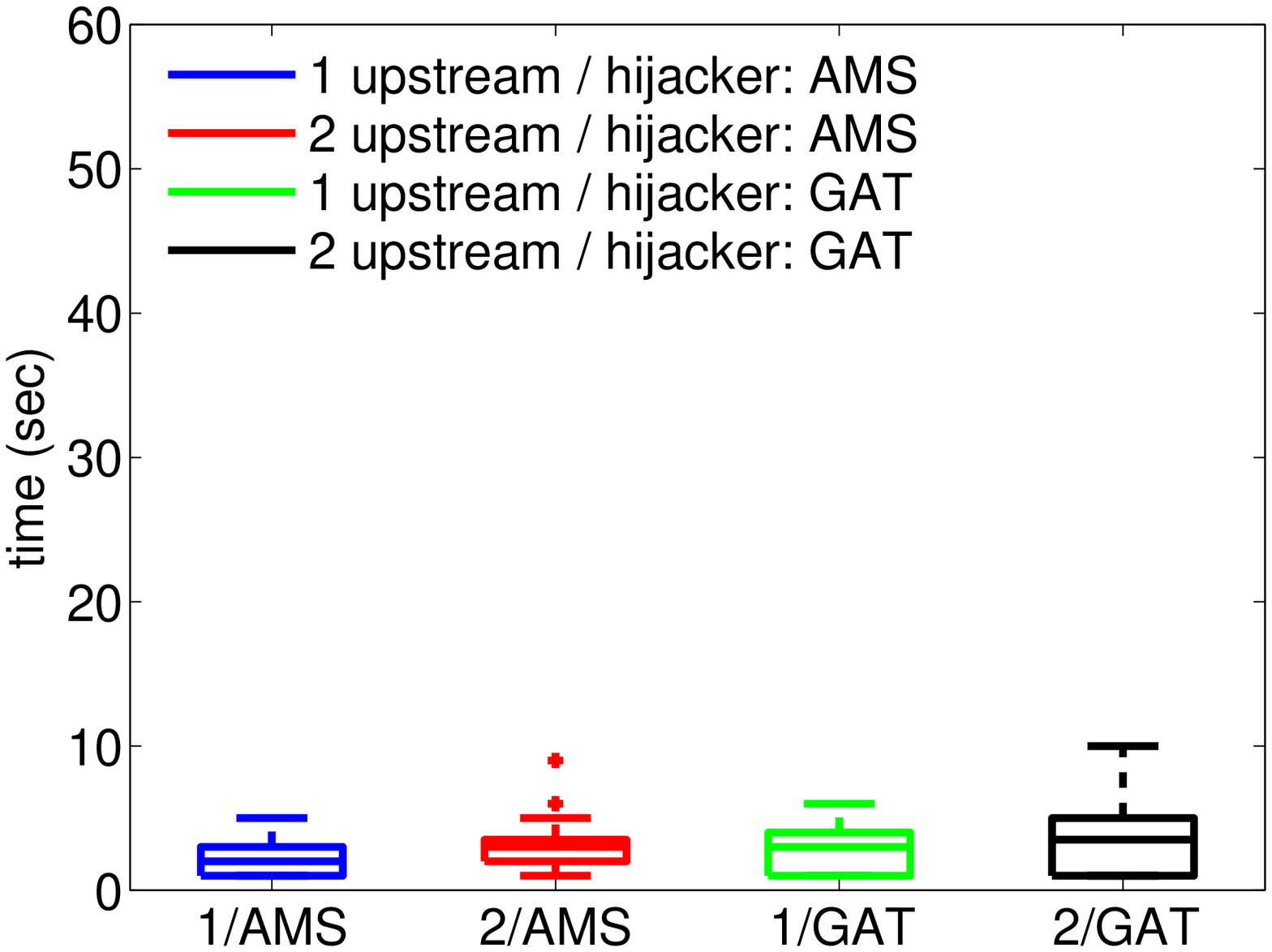}\label{fig:mboxplots-delay-vs-location-subprefix}}
\caption{Boxplots of ARTEMIS detection delay in scenarios with different \textit{hijacker location} and \textit{number of upstream providers} of the legitimate AS. Results for the cases of (a) prefix and (b) sub-prefix hijacking.}
\label{fig:detection-location}
\end{figure}

\begin{figure}
\subfigure[prefix hijacking]{\includegraphics[width=0.49\linewidth]{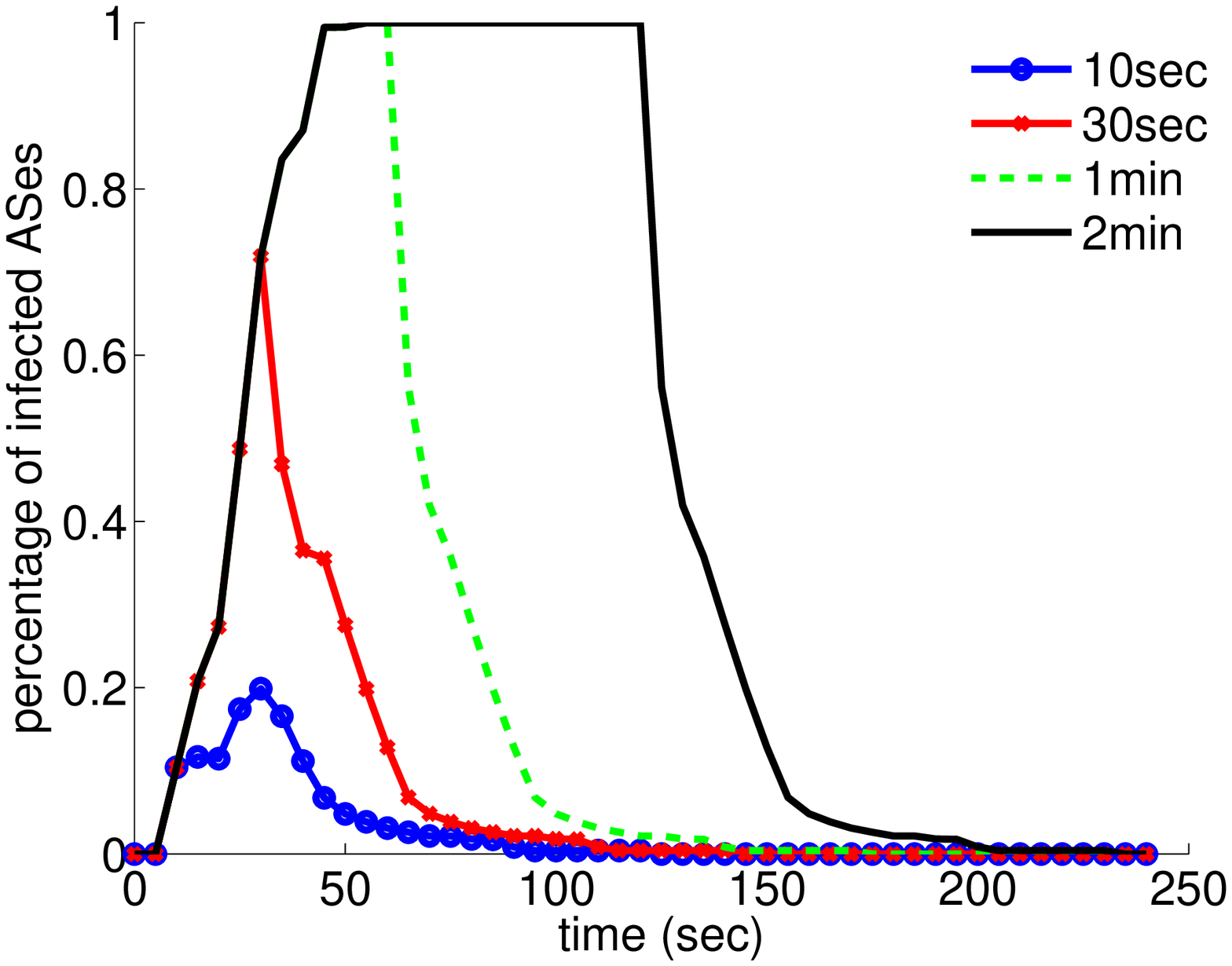}\label{fig:salvaged-AS-prefix}}
\subfigure[sub-prefix hijacking]{\includegraphics[width=0.49\linewidth]{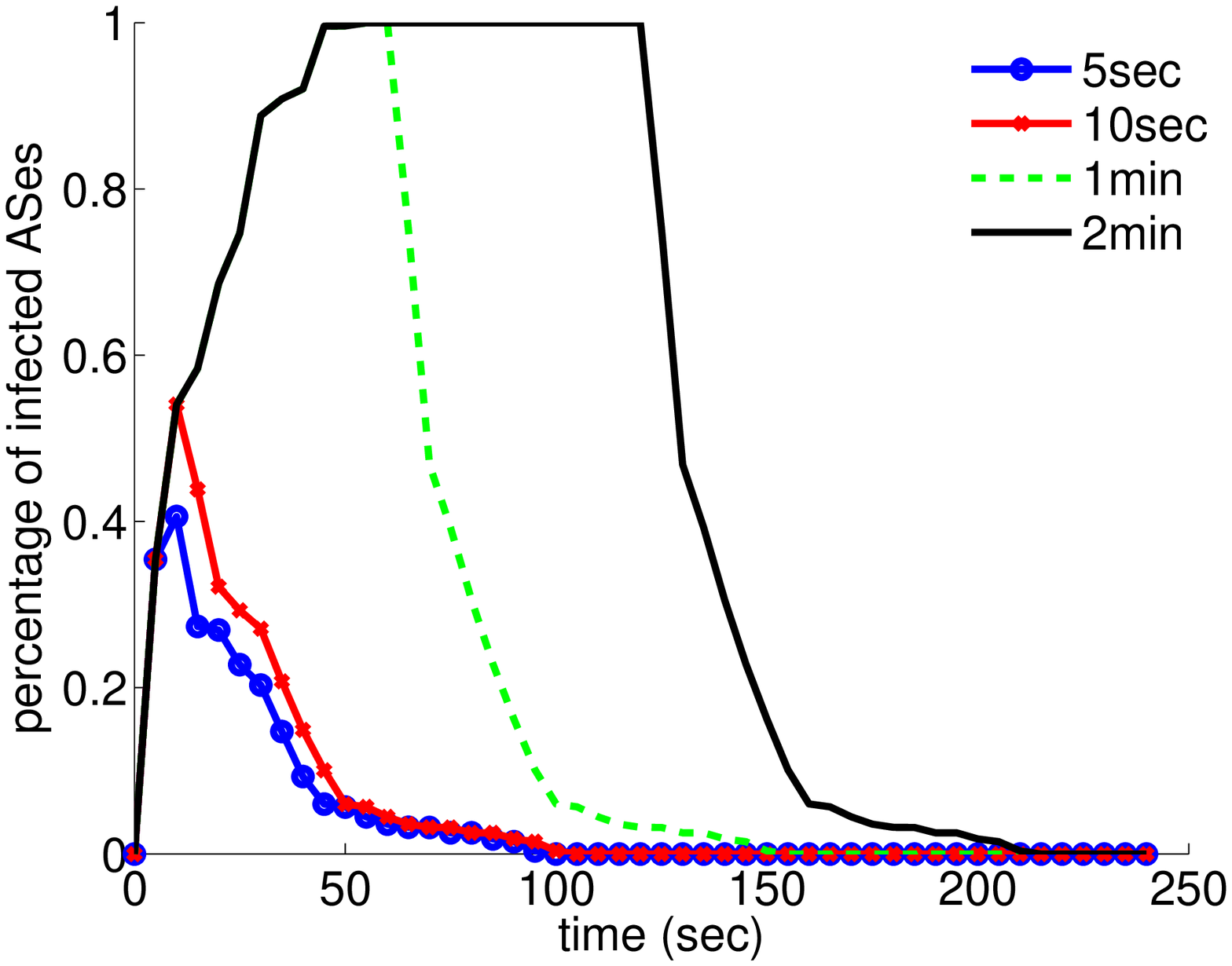}\label{fig:salvaged-AS-subprefix}}
\caption{Percentage of ``infected'' ASes (seen by the control-plane sources), i.e., ASes that route traffic to the hijacker, (y-axis) vs. the time since the hijacking has been launched (x-axis).}
\label{fig:percentage-mititgation}
\end{figure}

After presenting the hijacking detection efficiency, we study the gains of the automatic mitigation of ARTEMIS. Specifically, Fig.~\ref{fig:percentage-mititgation} shows the percentage of infected ASes in relation to the time since the hijacking has been launched. Each curve corresponds to a different ``mitigation start time'' $T_{start}$, i.e., the time between the hijacker's announcement and the de-aggregation. The two bottom lines\footnote{With blue and red; or, $10$sec and $30$sec in Fig.~\ref{fig:salvaged-AS-prefix}, and $5sec$ and $10sec$ in Fig.~\ref{fig:salvaged-AS-subprefix}.} correspond to the near real-time automatic mitigation with ARTEMIS (we selected representative scenarios; cf. Fig.~\ref{fig:detection-location}). The two top lines are assumed to correspond to a timely (but not real-time) mitigation, e.g., with manual actions.

As it can be seen, ARTEMIS can significantly decrease the impact of a hijacking. For instance, in scenarios where the detection delay of ARTEMIS is $10$sec, the fraction of infected ASes is $20\%$ and $50\%$ in the prefix and sub-prefix hijacking, respectively, while even a timely mitigation starting $1$min after the hijacking is not able to prevent the infection of all ASes. Moreover, with ARTEMIS the attack is completely mitigated in $T_{total}\leq2$min, whereas a mitigation that started after all ASes are infected (i.e., top two lines) needs around $1.5$min \textit{after the detection}, i.e., $T_{total} = T_{start}+1.5$min. 

This fast and effective mitigation that ARTEMIS can achieve, is particularly important for short hijacking attacks, whose frequency increases~\cite{Shi-Argus-IMC-2012}, and which can still cause serious problems~\cite{hijack-BitCoins}.

\section{Related Work}\label{sec:related}

\subsection{Detection of Prefix Hijacking}
Detection mechanisms can be classified based on the type of information they use for detecting prefix hijackings as: (i) control-plane, (ii) data-plane, and (iii) hybrid approaches.

Control-plane approaches~\cite{Chi-cyclops-CCR-2008,Lad-Phas-Usenix-2006,bgpmon} collect information, like BGP updates or tables, from route servers and/or looking glass servers (LGs), from which they detect incidents that can be caused by prefix hijackings. When, for example, a change in the origin-AS of a prefix, or a suspicious change in a route path, is observed, an alarm is raised. Data plane approaches~\cite{Zhang-Ispy-CCR-2008,Zheng-lightweight-hijacks-2007} use ping/traceroute measurements to detect a prefix hijacking. They continuously monitor the data plane connectivity of a prefix and raise an alarm for hijacking, when significant changes in the reachability of the prefix~\cite{Zhang-Ispy-CCR-2008} or in the paths leading to it~\cite{Zheng-lightweight-hijacks-2007}, are observed. A main shortcoming of data-plane mechanisms, is that a significant (minimum) number of active measurements is required to safely characterise an event as hijacking. Hence, these systems cannot be implemented in a light-weight fashion; and if deployed by every AS, they could introduce a large overhead~\cite{Shi-Argus-IMC-2012}. Finally, hybrid approaches ~\cite{Hu-accurate-hijacks-SP-2007,Shi-Argus-IMC-2012} combine control and data plane information to detect,  with higher accuracy~\cite{Shi-Argus-IMC-2012}, multiple types of prefix hijacking~\cite{Hu-accurate-hijacks-SP-2007,Shi-Argus-IMC-2012}.

Argus~\cite{Shi-Argus-IMC-2012} is the most recent among the aforementioned detection systems, and has few false positives/negatives, and near real-time detection. However, Argus is based only on BGPmon for control-plane information~\cite{bgpmon}, whereas ARTEMIS receives data from multiple sources (BGPmon, RIPE-RIS, BGPstream, Periscope), which leads to a faster detection in more than $60\%$ of the cases (as we observed in our experiments).

The main difference between ARTEMIS and previous detection mechanisms is that most of the previous approaches are \textit{notification systems}. They are designed to be operated by a third party and to monitor {\it all} the prefix the Internet. Upon detection of a suspicious event, they notify the involved ASes about a possible hijacking. This process has two shortcomings: (a) it yields many false positives, since suspicious events can usually be due to a number of legitimate reasons, like traffic engineering, anycast, congestion of the data-plane, etc.; and (b) it introduces significant delay, between the detection and mitigation of an event, as the network administrators of the involved AS need to be notified and then have to manually verify the incident. On the contrary, ARTEMIS is designed to detect hijacks against owned prefixes (for which the origin-AS information is known) and thus, overcomes the accuracy limitations, and eliminates the notification/verification delay.

Among previous works, only~\cite{Zhang-Ispy-CCR-2008} is designed for detection of hijacks against owned prefixes. However, it is a pure data-plane mechanism, which as discussed above, introduces significant overhead (especially, if deployed by many ASes). In contrast, ARTEMIS, which is a pure control-plane mechanism, can be deployed simultaneously in many ASes without significant overhead for the control-plane resources/tools (see discussion in Section~\ref{sec:detection}). 
At the time the first of these mechanisms were proposed, the capability of the available BGP feeds for providing real-time information was limited. However, currently there exist several state-of-the-art \textit{publicly available} control-plane sources/tools~\cite{bgpmon,ripe-ris-real-time,bgpstream,Giotsas-Periscope-PAM-2016} that enable pure control-plane mechanisms, like ARTEMIS, to detect a prefix hijacking event in near real-time (a few seconds~\cite{bgpmon,ripe-ris-real-time}, or minutes~\cite{Giotsas-Periscope-PAM-2016}, as we show in Section~\ref{sec:evaluation}). To our best knowledge ARTEMIS is the first approach to exploit the streaming interfaces of RIPE RIS~\cite{ripe-ris-real-time} and Periscope~\cite{Giotsas-Periscope-PAM-2016} for prefix hijacking detection.

\subsection{Measurement Studies}

Previous studies have taken measurements either over real hijacking incidents that happened in the Internet~\cite{Shi-Argus-IMC-2012, Lad2-Understanding-resiliency-hijacks-DSN-2007,Khare-concurrent-hijacks-IMC-2012} or through simulations~\cite{Zhang-Ispy-CCR-2008, Lad2-Understanding-resiliency-hijacks-DSN-2007,Zheng-lightweight-hijacks-2007,Zhang-practical-defenses-CoNext-2007}. While the former correspond to the  behavior of the Internet and capture the real effects of hijacking, they are limited to the investigation of a few known incidents, which do not span all possible cases. The latter are able to perform an investigation over a wider range of scenarios and study the effect of different parameters, but do not capture accurately  real-world effects, since the topology, routing, and policies of the Internet cannot be perfectly replicated in simulations. Moreover, due to the absence of the ground-truth, i.e., if a detected routing change is indeed a hijacking or not, previous studies refer to third party sources, e.g., the Route Origin Authorizations (ROA) or Internet Routing Registries (IRR), for the validation of their results. However, such information is usually incomplete and/or inaccurate~\cite{Shi-Argus-IMC-2012}, and, this might have an impact on the findings. Closer to our study is~\cite{Zhang-Ispy-CCR-2008} that tested its performance in self-triggered hijacks (for self-owned prefixes) in the Internet. Nevertheless, only few experiments (15 hijacks) were conducted, whereas in this paper we conduct a large number of experiments (spread over 4 weeks) with varying network parameters (location and connectivity of the hijacking/legitimate AS) and types of hijacks.

\section{Conclusion}\label{sec:conclusion}
We have presented ARTEMIS, a system for near real-time detection and automatic mitigation of BGP hijacking attacks. The evaluation with extensive real hijacking experiments, showed that ARTEMIS can detect hijacks in a few seconds, and completely mitigate them in less than $2$min.

In this initial implementation of ARTEMIS, we detect \textit{origin-AS} inconsistencies in route paths, and combat them using the \textit{prefix de-aggregation} method. Although not a panacea, prefix de-aggregation can be also effective for adjacency / policy~\cite{Shi-Argus-IMC-2012} or last-hop anomalies~\cite{Lad-Phas-Usenix-2006}, or even path interception attacks~\cite{caida-hijacks-project}. To extend ARTEMIS towards this direction, it suffices to modify only the detection algorithm; the monitoring and mitigation services can remain intact.

Finally, since the detection service of ARTEMIS is built on top (and, independently) of the monitoring service, the employed monitoring methodology and results (e.g., Fig.~\ref{fig:detection-mititgation}) are generic and can be useful in a number of application related to control-plane monitoring, e.g., to provide visibility to an AS of the impact of the routing changes it triggers (anycasting, traffic engineering, etc).

\bibliographystyle{ieeetr}

\end{document}